\documentclass[namedreferences]{kluwer}[1998/02/11]
\usepackage[dvips]{epsfig}

\def\lesssim{\mathrel{\hbox{\rlap{\hbox{\lower4pt\hbox{$\sim$}}}\hbox{$<$}}}}
\def\gtrsim{\mathrel{\hbox{\rlap{\hbox{\lower4pt\hbox{$\sim$}}}\hbox{$>$}}}}
\let\la=\lesssim
\let\ga=\gtrsim
\def\arcsec {\hbox{$^{\prime\prime}$}}

\newcommand{\farcs}{\hbox{$.\!\!^{\prime\prime}$}}

\newcommand{\HII}{H$\;${\small\rm II}\relax}
\newcommand{\Ha}{H$\alpha$}

\newcommand{\etal}{{\em et al.}\relax}
\newcommand{\z}{$z$}

\newcommand{\msun}{M$_\odot$}

\newcommand{\e}[1]{10^{#1}}
\newcommand{\ngc}{NGC~}

\newcommand{\pI}{Paper~I}
\newcommand{\pII}{Paper~II}
\newcommand{\pIII}{Paper~III}

\newcommand{\aj}[1]{{\it Astron. J.} {\bf #1}}
\newcommand{\apj}[1]{{\it Astrophys. J.} {\bf #1}}
\newcommand{\aap}[1]{{\it Astron. Astrophys.} {\bf #1}}
\newcommand{\mnras}[1]{{\it MNRAS} {\bf #1}}

\begin{document}     
\begin{article}
\begin{opening}

\title{Extraplanar Dust in Spiral Galaxies: Observations and 
	Implications}

\author{J. Christopher \surname{Howk} \email{howk@pha.jhu.edu}}
\institute{The Johns Hopkins University, Dept. of Physics \& Astronomy,
\\Baltimore, MD 21218, USA}

\runningtitle{Extraplanar Dust}
\runningauthor{J.C. Howk}

\begin{ao}
\\The Johns Hopkins Univ., Dept. of Physics \& Astronomy \\
3400 N. Charles St., Baltimore, MD 21218,  USA \\
e-mail: howk@pha.jhu.edu\\
\end{ao}

\begin{abstract}

Recent optical and submillimeter observations have begun to probe the
existence of dust grains in the halos of spiral galaxies.  I review
our own work in this area which employs high-resolution optical images
of edge-on spiral galaxies to trace high-\z\ dust in absorption
against the background stellar light of the galaxies.  We have found
that a substantial fraction of such galaxies ($>50\%$) show extensive
webs of dust-bearing clouds to heights $z>2$ kpc.  Extraplanar dust in
galaxies is statistically correlated with extraplanar diffuse ionized
gas, though there is no evidence for a direct, physical relationship
between these two phases of the high-\z\ interstellar medium.  The
dense high-z clouds individually have masses estimated to be $\ga
\e{5} - \e{6}$ \msun.  The detailed properties of the observed dust
structures suggest the clouds seen in our images may represent the
dense phase of a multiphase ISM at high-\z.  Such dense clouds can
have an important effect on the observed light distribution in spiral
galaxies.  I discuss the effects such high-\z\ dust can have on
quantitative measures of the vertical structure of stars and ionized
gas in edge-on systems.
\end{abstract}

\end{opening}  

\section{Introduction}

Dust in the thin disks of spiral galaxies gives rise to substantial
extinction, and the effects of this extinction dominate the optical
appearance of spirals when viewed edge-on.  Recent work in the optical
and submillimeter wavebands has shown that dust grains are not only to
be found in the thin disks, but also far into the thick disks and
halos of spiral galaxies (Zaritsky 1994; Alton \etal\ 1998; Block
\etal\ 1999).  The energy input associated with star formation in
spiral galaxies can expel dust grains and gas from their disks to the
surrounding interstellar halos. The expulsion of grains from the disks
of galaxies can be driven by violent hydrodynamical processes (e.g.,
Norman \& Ikeuchi 1989) or more quiescent processes such as radiation
pressure (e.g., Davies \etal\ 1998).  This expulsion of gas and the
heavy element-rich dust grains from the disks of galaxies may have
important implications for the chemical evolution of intracluster or
intergalactic media (Wiebe
\etal\ 1999).  Furthermore, if the dust is present in great enough
quantities it can have important effects on the perceived distribution
of light.  But, until recently, little was known of the dust content
of the interstellar medium (ISM) far above the planes of galaxies.

\begin{figure}
\centerline{\epsfig{file=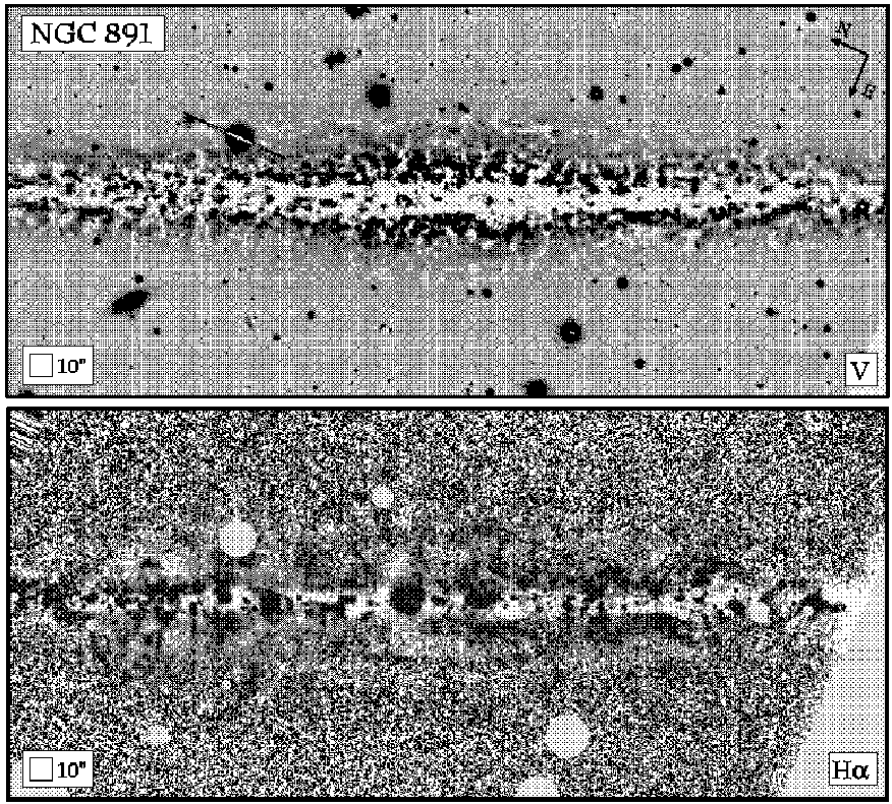, width=\maxfloatwidth}}
\caption{WIYN unsharp-masked V-band (top) and \Ha\ (bottom) images of
\ngc 891 (adopted from Howk \& Savage 1999b).  The 10\arcsec\ box
corresponds to 450 pc at the distance of \ngc 891 (9.5 Mpc).  The
unsharp masked versions of the data were derived using a smoothing
Gaussian with FWHM$\, = 35$ pixels ($6\farcs9$ or 310 pc).  The
seeing-limited resolution of these images are $\sim0\farcs8$ (36 pc)
and $0\farcs9$ (40 pc) for the V-band and \Ha\ data, respectively.
The grayscale in both images is inverted so that dark areas correspond
to brighter emission.}
\label{fig:n891}
\end{figure}

In Howk \& Savage (1997, hereafter \pI), we presented high-resolution
optical images of the nearby edge-on spiral galaxy
\ngc 891 taken with the WIYN 3.5-m telescope.\footnote{The WIYN
Observatory is a joint facility of the University of
Wisconsin-Madison, Indiana University, Yale University, and the
National Optical Astronomy Observatories.}  These images reveal an
extensive web of high-\z\ dust structures seen in absorption against
the background stellar light of the bulge and thick stellar disk.
Figure \ref{fig:n891} shows a deep WIYN V-band image of \ngc 891 (top
panel; adapted from Howk \& Savage 1999b, hereafter \pIII).  The
absorbing dust structures are seen all along the imaged portion of the
galaxy to heights approaching $z \sim 2$ kpc.  These dust-bearing
clouds are optically thick and each contain significant amounts of mass ($\ga
\e{5}-\e{6}$ \msun) and very likely contain molecular material.

In this contribution I review our current understanding of the
physical and statistical properties of such high-\z\ clouds in spiral
galaxies (\S \ref{sec:dust}).  I also discuss the effects of these
opaque high-\z\ dust clouds on quantitative measures of galaxy
structure (\S \ref{sec:light}).

\section{Extraplanar Dust in Normal Spiral Galaxies}
\label{sec:dust}

\subsection{Observations}

In this section I summarize the observational state of our knowledge
of high-\z\ dust-bearing clouds like those seen in Figure
\ref{fig:n891} and discuss briefly our theoretical understanding of
these structures.  Our current observational understanding of
extraplanar dust in normal spiral galaxies can be briefly summarized
in three primary conclusions:

{\bf 1) The observable dust-bearing clouds are clumpy and highly
structured and can be viewed to heights $z \la 2$ kpc; they have large
opacities and likely trace large amounts of mass.}

Figure \ref{fig:n891} shows that the distribution of high-\z\
absorbing material in \ngc 891 is complex.  In a few cases the dust
structures seem obviously to be tracing disk-halo outflows or
interactions.  However, the vast majority of the observable dust
clouds show no immediate morphological connection with canonical
chimney or superbubble phenomena (see \pI).

The absorption due to this high-\z\ dust is significant, with each of
the hundreds of dust-laden structures providing $A_V \ga 1$ mag.
These dust-bearing clouds, if they have Galactic dust-to-gas ratios,
have masses $\ga \e{5}-\e{6}$ \msun, similar to the Galactic giant
molecular clouds.  As an ensemble the clouds likely contain $\ga
\e{8}$ \msun, or approximately the same amount of mass as the diffuse
ionized gas (DIG; Dettmar 1990).  The maintenance of such an extended
layer of dusty material requires a significant amount of energy: the
gravitational potential energy for each cloud is $>\e{52}$ ergs (\pI).
Though our data are deep enough to reveal such clouds, there is a real
paucity of structured dust-laden clouds at heights $z\ga 2$ kpc
(\pIII).

{\bf 2) Extraplanar dust-bearing clouds are a common feature of spiral
galaxies, and their existence is statistically correlated with the
presence of extraplanar diffuse ionized gas.}

In Howk \& Savage (1999a, hereafter \pII) we presented short
observations of all the {\em truly} edge-on, massive northern spiral
galaxies within 25 Mpc.  Table \ref{tab:statistics} lists the seven
such galaxies, and indicates whether they show evidence for
extraplanar dust.  Also indicated is the presence or absence of
extraplanar DIG.  The galaxies in this table are listed in order of
decreasing far-infrared luminosity per unit disk area, a rough
indicator of the star-formation rate per unit area.

\begin{figure}
\centerline{\epsfig{file=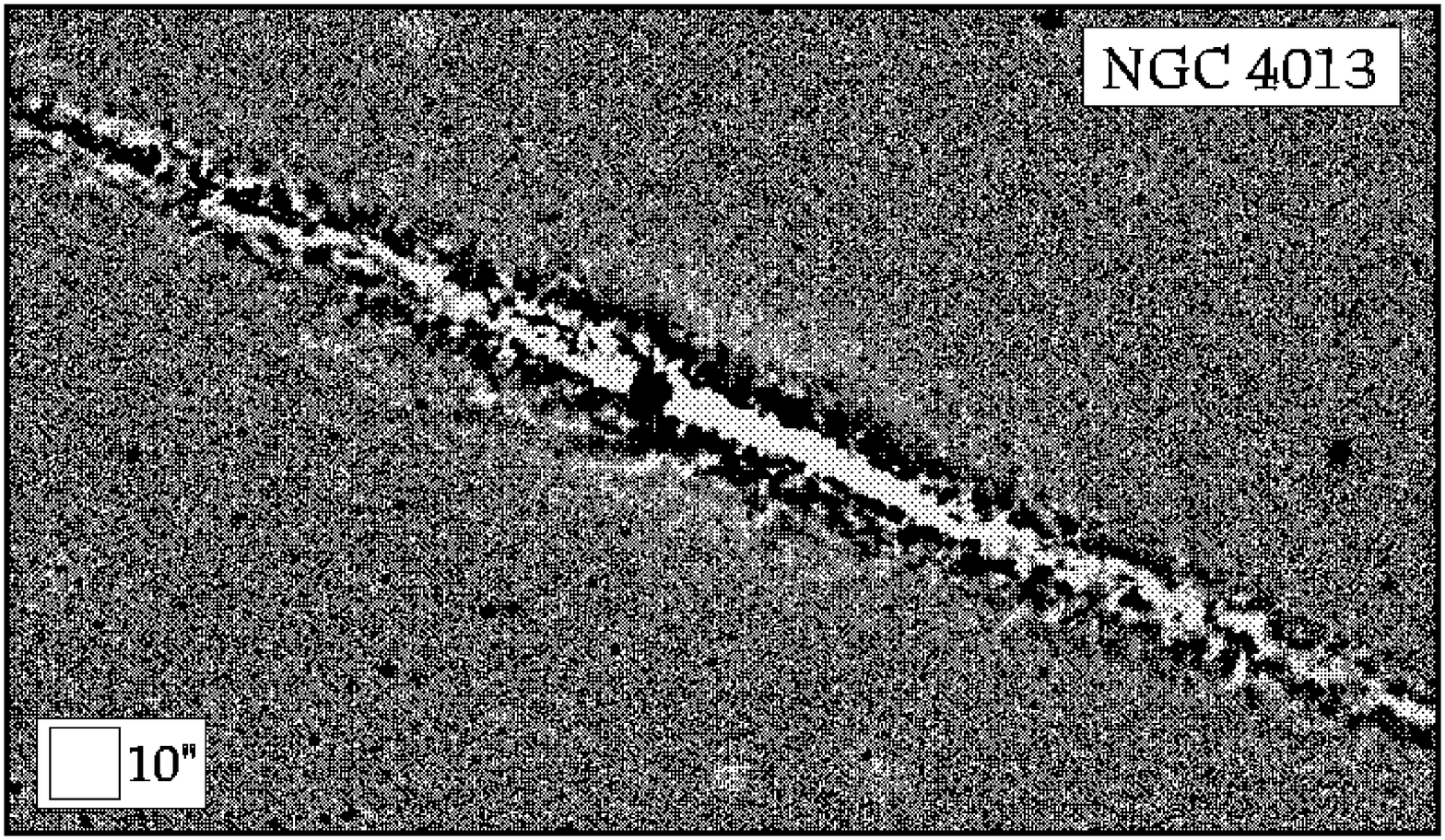, width=4.35in}}
\caption{WIYN unsharp-masked V-band image of \ngc 4013.
At the distance of this galaxy ($D\sim17$ Mpc), the 10\arcsec\ box
corresponds to $\sim825$ pc on a side.  The seeing limited resolution
of this image is $0\farcs5$ (41 pc).  These data are discussed more
fully in Howk \& Savage (1999a).}
\label{fig:n4013}
\end{figure}

Though there are a small number of truly edge-on galaxies in the
survey volume, Table \ref{tab:statistics} shows a large fraction of
local universe spirals contain extraplanar dust.  Furthermore, there
is a one-to-one correlation between the presence (or absence) of
extraplanar dust and the presence (or absence) of extraplanar DIG.
Figure \ref{fig:n4013} shows a WIYN unsharp-masked V-band image of
\ngc 4013 (\pII).  The dust structures in this system have similar 
properties to those seen in \ngc 891 (see \pII).  Given the short
exposure times required to identify such high-\z\ absorbing
structures, using dust as a tracer of interstellar matter in galaxy
halos is much more efficient than \Ha\ imaging.


\newcommand{\wyes}{\Large $\bullet$ \normalsize}
\newcommand{\wno}{\Large $\circ$ \normalsize}
\small
\begin{table}
\begin{tabular*}{\maxfloatwidth}{lcccccc}
\hline
Galaxy & L$_{FIR}$/D$^2_{25}$ & High-z & & High-z & & DIG \\ &
(10$^{40}$ erg/s kpc$^2$) & Dust$^a$ & & DIG$^b$ & & Ref.\\
\hline
NGC 891  & 3.3  & \wyes & &  \wyes  & &  1,2 \\
NGC 4013 & 2.6  & \wyes & &  \wyes  & &  3   \\
NGC 4302 & $\la$2.3 	        
                & \wyes & &  \wyes  & &  3  \\
NGC 3628 & 1.8  & \wyes & &  \wyes  & &  4   \\
NGC 5907 & 0.8  & \wno  & &  \wno   & &  3    \\ 
NGC 4565 & 0.3  & \wno  & &  \wno   & &  5    \\
NGC 4217 &  $<$0.12 		       
                & \wyes & &  \wyes  & &  3    \\
\hline 
\multicolumn{7}{l}{\scriptsize $a$ -- $\bullet$ denotes galaxies that exhibit 
	high-$z$ dust in WIYN images; $\circ$ denotes }\\ 
\multicolumn{7}{l}{\hspace{0.5 cm} \scriptsize galaxies that do not exhibit 
	high-$z$ dust in WIYN images.}\\
\multicolumn{7}{l}{\scriptsize $b$ -- $\bullet$ denotes galaxies with observable 
	high-$z$ DIG; $\circ$ denotes galaxies }\\
\multicolumn{7}{l}{\hspace{0.5 cm} \scriptsize  for which H$\alpha$ searches have 
	not shown detectable high-$z$ DIG.}\\
\multicolumn{7}{l}{\scriptsize REFERENCES: (1) Dettmar (1990); 
	(2) Rand \etal\ (1990); (3) Rand (1996);}\\ 
\multicolumn{7}{l}{\scriptsize (4) Fabbiano \etal\ (1990); 
	(5) Rand \etal\ (1992)}\\
\end{tabular*}
\caption[]{Dust and DIG Properties for Edge-On Galaxy Sample}
\label{tab:statistics}
\end{table}

\normalsize

{\bf 3) There is NO evidence for a direct physical relationship
between the dust-bearing clouds and the DIG seen in galaxy halos.}

Though there is a striking statistical correlation between the
presence of high-\z\ dust and DIG in spiral galaxies (\pII), the
morphology of the dust and the DIG are extremely different.  The \Ha\
emission from the halo of \ngc 4013 (Rand 1996) is localized to only a
few patches, and seems to be more smoothly distributed within those
patches, whereas the high-\z\ dust absorption is visible along most of
the length of the galaxy and is highly structured.

A more detailed, direct comparison between the \Ha\ and dust
morphologies for \ngc 891 is given in \pIII. Figure \ref{fig:n891}
shows an unsharp-masked version of the WIYN \Ha\ image immediately
below the V-band unsharp-masked image.  Comparing the two images in
Figure \ref{fig:n891}, one is left with the (correct) impression that
the \Ha\ distribution is much smoother with less pronounced, and
fewer, filamentary structures.

\subsection{Interpretation}

The comparison of the \Ha\ and broadband data for \ngc 891 (\pIII)
suggest that we are observing two distinct ``phases'' of the ISM with
these two probes of thick disk matter.  The \Ha\ images trace an
ionized, low-density medium, while the dust structures seen in
absorption are tracing a dense, neutral phase of the high-\z\ ISM.  In
standard models of thermal phase equilibrium for disk of the Milky
Way, a dense phase of the ISM is present only if the pressure of the
medium is high enough (e.g., Wolfire \etal\ 1995).  Rough calculations
of the pressure distribution of the DIG and hot ISM in \ngc 891
suggest that these media provide a sufficient external pressure to
allow for the existence of a dense neutral phase (see \pII).  That
virtually no clouds are observable at heights in excess of $z\sim2$
kpc may imply that the pressure is insufficient to support the dense
clouds above this height.  If this is the case, there may very well be
unobserved dust at heights in excess of 2 kpc (our images are only
sensitive to clouds that are overdense relative to their
surroundings).  In general we do not expect a cold, neutral phase of
the ISM at high-\z\ to be directly associated with the extraplanar
DIG.

The interpretation that the observed dust clouds represent the dense
phase of a multiphase medium at high-\z\ is also supported by two
recent observational results.  Garc\'{\i}a-Burillo \etal\ (1999) have
presented high-resolution CO maps of the inner portion of
\ngc 4013 showing extraplanar CO features far above the plane of the 
galaxy. {\em Many of these CO features are directly associated with
high-\z\ dust structures observable in our WIYN images.}  The presence
of a dense phase of the thick disk ISM is also implied by our own
discovery of \HII\ regions far above the planes of \ngc 891 and \ngc
4013 (see \pIII).  In \ngc 891 these candidate nebulae are associated
with faint, blue continuum sources at heights $0.5 \la z \la 2.0$ kpc.
These objects likely represent young stellar associations formed far
above the plane of the galaxy, perhaps in the dense clouds visible in
our WIYN images.

\section{Extraplanar Dust and the Vertical Light Distribution
in Spiral Galaxies}
\label{sec:light}

The presence of dense dust-laden (molecular) clouds in the thick
interstellar disks of spiral galaxies affects the perceived vertical
distribution of light in these systems.  The absorption due to these
clouds will not only affect the perceived distribution of stellar
light, but will also affect light produced by the high-\z\ gas layers
(e.g., \Ha\ or X-ray emission).  In \pIII\ we have shown that much of
the structure perceived in the \Ha\ emission from the DIG at heights
$z\la1.0$ kpc from the midplane of \ngc 891 is caused by the patchy
absorption of these dense clouds.

\begin{figure}
\centerline{\epsfig{file=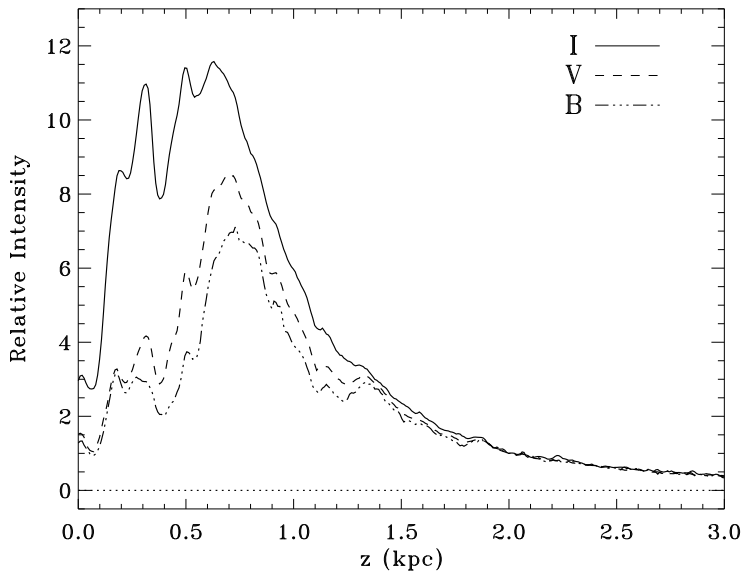, width=4.1in}}
\caption{Vertical light profile of \ngc 891 near the center
of the galaxy in the B,V, and I wavebands.  Each band has been
normalized to its intensity at $z=2$ kpc.  The curves represent an
average over only 45 pc in the radial direction (just larger than the
FWHM of the seeing disk).  }
\label{fig:cuts}
\end{figure}

Figure \ref{fig:cuts} shows the vertical light distribution for a
region near the center of \ngc 891, averaged over 45 pc in the radial
direction and normalized to the intensity of each bandpass at $z=2$
kpc.  The impact of discrete absorbing clouds can be seen in these
vertical cuts.  For example, quite prominent dust structures are
intercepted at $z \sim 1.2$ kpc.  One can see by comparing the B- and
I-band curves in Figure \ref{fig:cuts} that the distribution of light
at heights $z\la 0.75$ kpc is severely affected by the presence of
extraplanar interstellar dust.  Fits to the light distribution at
heights less than this will be dominated by dust absorption.  Figure
\ref{fig:cuts} also shows that \ngc 891 gets {\em bluer} with distance
from the midplane to a height of $z \sim 1.9$ kpc, indicating that
large amounts of dust are present at great heights.

Most studies of the structure of nearby galaxies employ images taken
in much poorer seeing than ours.  This has the effect of washing out
the discreet, small-scale dust structures.  Typical high-\z\ dust
features in the thick disk of \ngc 891 have minor axis sizes of order
100 pc ($\la2$\arcsec).  Images taken in seeing $\ga 1\farcs5$ show
relatively little direct evidence for the widespread, complicated
distribution of dust seen in our higher resolution data.

The high-\z\ dust structures have two important effects on the
perceived vertical distribution of light.  First, it tends to flatten
the observed distribution of emission, making the scale height of the
light seem {\em larger} than the true scale height.  Second, since the
dust is more prevalent at low \z -heights, it can also cause two
distinct vertical components (e.g., thin and thick disks) to appear as
one in lower resolution data.  Single exponential fits to smoothed
versions of our WIYN B-band data (simulating seeing of $1\farcs5$),
for example, yield a scale height $h_z = 1.05\pm0.04$ kpc with a
linear correlation coefficient $r=0.995$ (fitting only the light above
$z = 1.0$ kpc in log space).  The true distribution of emission,
however, is likely similar to the I-band, where two components are
clearly seen: a thin disk whose scale height is uncertain because of
the dust (Xilouris \etal\ 1998 derive $h_z \approx 0.3$ kpc) and a
thick disk with scale height of order $h_z \sim 1.7$ kpc (Morrison
\etal\ 1997).

It is likely that extraplanar dust significantly affects the perceived
distribution of light away from the midplanes of spiral galaxies.

The effects of extraplanar dust on the perceived light distribution in
spiral galaxies can be extremely important for studies attempting to
derive fundamental morphological properties of these systems.  This is
true for studies of the starlight as well as studies of emission from
the ISM above the planes of spirals, including \Ha\ and soft X-ray
emission.

\acknowledgements

My thanks to the Center for Astrophysical Studies at The Johns Hopkins
University for travel assistance.

\end{article}

\begin{thebibliography}{}

\bibitem[]{} Alton, P.B., Bianchi, S., Rand, R.J., Xilouris, E.M., Davies, 
J.I., \& Trewhella, M.: 1998, \apj{507} L125

\bibitem[]{} Block, D.L., Stockton, A., Elmegreen, B.G., \& Willis, J.:
1999, \apj{522}, 25

\bibitem[]{} Davies, J.I., Alton, P., Bianchi, S., \& Trewhella, M.: 1998, 
\mnras{300}, 1006

\bibitem[]{} Dettmar, R.-J.: 1990, \aap{232}, L15

\bibitem[]{} Fabbiano, G., Heckman, T.M., \& Keel, W.C.: 1990, \apj{355}, 442

\bibitem[]{} Garc\'{\i}a-Burillo, S.,  Combes, F., \& Neri, R.: 1999,
\aap{343}, 740

\bibitem[]{} Howk, J.C., \& Savage, B.D.: 1997, \aj{114}, 2463 (\pI)

\bibitem[]{} Howk, J.C., \& Savage, B.D.: 1999a, \aj{117}, 2077 (\pII)

\bibitem[]{} Howk, J.C., \& Savage, B.D.: 1999b, \aj{} submitted. (\pIII)

\bibitem[]{} Morrison, H.L., Miller, E.D., Harding, P., Stinebring,
D.R., \& Boroson, T.A.: 1997, \aj{113}, 2061

\bibitem[]{} Norman, C.A. \& Ikeuchi, S.: 1989, \apj{345}, 372

\bibitem[]{} Rand, R.J.: 1996, \apj{462}, 712

\bibitem[]{} Rand, R.J., Kulkarni, S.R. \& Hester, J.J.: 1990, \apj{352}, L1

\bibitem[]{} Rand, R.J., Kulkarni, S.R. \& Hester, J.J.: 1992, \apj{396}, 97


\bibitem[]{} Wiebe, D.S., Shustov, B.M., \& Tutokov, A.V.: 1999, 
\aap{345}, 93

\bibitem[]{} Wolfire, M.G., Hollenbach, D., McKee, C.F., Tielens, 
A.G.G.M., \& Bakes, E.L.O.: 1995, \apj{443}, 152

\bibitem[]{} Xilouris, E.M., Alton, P.B., Davies, J.I., Kylafis, N.D., 
Papamastorakis, J., \& Trewhella, M.: 1998, \aap{331}, 894

\bibitem[]{} Zaritsky, D.: 1994, \aj{108}, 1619


\end{thebibliography}
\end{document}